\documentclass[twocolumn,nofootinbib]{revtex4-1}
\usepackage{latexsym,epsfig,amssymb, amsmath,nicefrac}

\usepackage{amsfonts,amsthm} 
\usepackage{bm,bbm}
\usepackage{graphicx}
\usepackage{esint}
\usepackage{color}
\usepackage{braket}

\newcommand{\roughly}[1]{\mathrel{\raise.3ex\hbox{$#1$\kern-0.85em
\lower1ex\hbox{$\sim$}}}}
\newcommand{\lsim}{\roughly<}
\newcommand{\gsim}{\roughly>}

\def\ssT{{\scriptscriptstyle T}}

\newcommand{\be}{\begin{equation}}
\newcommand{\ee}{\end{equation}}
\newcommand{\ba}{\begin{eqnarray}}
\newcommand{\ea}{\end{eqnarray}}
\def\bs{\begin{subequations}}
\def\es{\end{subequations}}

\def\cO{{\cal O}}

\begin{document}

\title{{\Large Inflation by Alignment}}

\author{C.P.~Burgess${}^{1,2,3}$ and Diederik Roest${}^4$}

\affiliation{{}$^1$ PH\,-TH Division, CERN, CH-1211, Gen\`eve 23, Suisse.}

\affiliation{{}$^2$ Department of Physics \& Astronomy, McMaster University\\ \qquad 1280 Main Street West, Hamilton ON, Canada.}

\affiliation{{}$^3$ Perimeter Institute for Theoretical Physics\\
\qquad 31 Caroline Street North, Waterloo ON, Canada.}

\affiliation{{}$^4$Van Swinderen Institute for Particle Physics and Gravity, University of Groningen, \\ Nijenborgh 4, 9747 AG Groningen, The Netherlands, d.roest@rug.nl}

\begin{abstract}
Pseudo-Goldstone bosons (pGBs) can provide technically natural inflatons, as has been comparatively well-explored in the simplest axion examples. Although inflationary success requires trans-Planckian decay constants, $f \gsim M_p$, several mechanisms have been proposed to obtain this, relying on { (mis-)}alignments between potential and kinetic energies in multiple-field models. We extend these mechanisms to a broader class of inflationary models, including in particular the exponential potentials that arise for pGB potentials based on noncompact groups (and so which might apply to moduli in an extra-dimensional setting). The resulting potentials provide natural large-field inflationary models and can predict a larger primordial tensor signal than is true for simpler single-field versions of these models. In so doing we provide a unified treatment of several alignment mechanisms, showing how each emerges as a limit of the more general setup.
\end{abstract}

\maketitle

\smallskip

\section{Introduction}

Inflationary models continue to provide a good description of the primordial fluctuations whose imprint is seen in precision measurements \cite{Planck} of the Cosmic Microwave Background (CMB). Because slow-roll inflation requires the existence of very light scalar fields, pseudo-Goldstone Bosons (pGBs) \cite{pGB,pGBrev} provide among the best-motivated inflationary models, at least from an ultraviolet (UV) perspective.

pGB models generically involve two energy scales: the decay constant, $f$, and the scale, $\Lambda$, of the potential. For inflationary purposes these must be chosen to obtain an inflationary slow-roll, and this is typically accomplished in either of two ways. In the case of a  pGB corresponding to a compact symmetry, slow-roll occurs when the relevant field $\phi$ takes generic values and so slow-roll relies on achieving $f \gsim M_p$ \cite{NatInf}. In constrast, in the case of a noncompact symmetry, slow-roll relies instead on the pGB field $\phi \gg f$, and not on the value of $f$ itself \cite{GL,LFInf}. In either case, obtaining observably large primordial tensor modes generically also requires $f \gsim M_p$ \cite{Lyth}.

Both kinds of pGB models can arise plausibly from UV completions, like string theory \cite{StrInf}. Within a string context the compact approach usually arises with $\phi$ an axion, and so with $f$ of order the string scale, $f \sim M_s \ll M_p$, where the inequality expresses how the string scale is lower than the Planck scale for weakly coupled strings. Noncompact models often arise when $\phi$ describes a modulus, such as an extra-dimensional shape parameter, for which $f \sim M_p$.

An important puzzle asks how larger values of $f$ can be obtained, and any successful enlargement of $f$ comes with a bonus: if $f \gg M_p$ then a small-field Taylor expansion of the potential (which relies on $\phi \ll f$) can be consistent with large-field inflationary predictions, such as observable tensor modes (which rely on $\phi \gsim M_p$). The hierarchy $M_p \lsim \phi \ll f$ brings the predictions of $\phi^2$ inflation \cite{PhiSqInf} to a wide class of pGB models \cite{Nem1}.

Two closely related proposals for obtaining large $f$ exist in cases where there is more than one pGB (which is fairly common for string vacua). In this case the low-energy scalar dynamics is controlled by the effective lagrangian
\begin{align}
   \mathcal{L} = \sqrt{-g} \left[ R - \tfrac12 K_{ij}(\phi) \, \partial \phi^i \partial \phi^j - V(\phi) \right] \,, \label{lattice}
\end{align}
where $V$ is the scalar potential and $K_{ij}(\phi)$ is the target-space metric. Although it is always possible to arrange $K_{ij}(\phi_0) = \delta_{ij}$ at a specific field point, $\phi_0$, in general this cannot be done everywhere unless the target-space geometry is flat (which is sometimes the case). In what follows we assume $K_{ij}$ can be transformed into $\delta_{ij}$, either because the target space is flat or because inflation takes place over a small enough region of it.

The mechanisms for obtaining large $f$ rely on a mismatch between the directions in field space that are preferred by the kinetic terms and the scalar potential. For example the basis which brings kinetic terms to diagonal form defines one set of directions. However, it can happen that the potential is simpler in a different basis, such as for axion models where strong interactions generate simple potentials for specific directions, which we call the {\em lattice} basis:
\be
 V = \sum_i \Lambda_i^4 \left( 1 - \cos \theta_i \right) \,.
\ee
Written in a basis with canonical kinetic terms --- the {\em kinetic} basis --- each term in the scalar potential contains a linear combination of the fields,
\begin{align}
   \mathcal{L} = \sqrt{-g} \left\{ \tfrac{1}{2} M_p^2 \, R - \tfrac12 (\partial \phi^i)^2 - \sum_i \Lambda_i^4 \left[ 1 - \cos( Q^i{}_j \phi^j ) \right] \right\} \,, \label{kinetic}
\end{align}
where the matrix $Q$ satisfies $Q^\ssT K Q = I$. Alternatively, in the lattice basis, the complications lie in the non-trivial kinetic term.

The simplest way employing multiple fields to enhance $f$ is the assisted-inflation mechanism \cite{assisted}, such as used in $N$-flation \cite{N-flation} and related models \cite{AA1}. Here there are $N$ axions and the $N\times N$ matrix $Q$ is taken to be diagonal with eigenvalues $1/f_i$. If the slow-roll during inflation is along a `diagonal' direction in moduli space, then the effective decay constant for the inflaton is given by the Pythagorean sum $(f_1^2 + \ldots + f_N^2)^{1/2}$, which, when all the individual constants are equal: $f_i = f \; \forall i$, simplifies to  $\sqrt{N} \; f$.

Alignment can lead to even more enhancement in more complicated situations, such as:
 \begin{itemize}
  \item
 For a large number of fields $N$ based on a random-matrix-theory approach, Bachlechner {\em et al} \cite{Bachlechner1} point out that the eigenvectors of the kinetic matrix are overwhelming likely to align with the diagonals in moduli space. Intuitively this phenomenon, known as {\it kinetic alignment}, occurs because the $2^N$ diagonals far outnumber the $N$ basis vectors. Compared to N-flation, where the length of all diagonals is set by the Pythagorean sum,
this leads to a splitting of the diagonal lengths ranging from $\sqrt{N} f_{\rm min}$ to $\sqrt{N} f_{\rm max}$, where $f_{\rm min(max)}$ is the {\em smallest (largest)} of the separate decay constants. Inflation benefits by taking place along the longest diagonal.
\item
 In the case of two fields, Kim, Nilles and Peloso \cite{KNP} point out that super-Planckian decay constants can arise when the directions defined by the scalar potential in \eqref{kinetic} are sufficiently aligned. When this happens the `other' direction becomes sufficiently flat to allow inflation to proceed. We refer to this as {\it lattice alignment}. This can be re-expressed as an increase in $f$, which even for two fields can be very high.
\end{itemize}
One of the points of this note is to demonstrate explicitly, in the simplest case of two fields, how these types of alignment are related and how they can be usefully combined. To this end, we will first review the case of alignment in natural inflation, and subsequently generalize to other functions based on non-compact pGBs.

\smallskip

\noindent
{\bf Note added:} upon completetion of this manuscript, we became aware of the preprint \cite{Bachlechner2} discussing related isses.

\section{Compact case}

Here we recap natural inflation \cite{NatInf} for two $U(1)$ axion fields whose potentials arise from non-perturbative symmetry-breaking effects. Our goal is to highlight the role played by the two alignment mechanisms described above.

Once tuned to vanish at its minimum, the scalar potential can be compactly written as
  \begin{eqnarray}
  V &=& \Lambda^4 \left[ 1- \tfrac12 \cos(\vec\alpha \cdot \vec\phi) -  \tfrac12 \cos(\vec\beta \cdot \vec\phi) \right] \nonumber \\
  &=& \Lambda^4 \left[ \sin^2 \left(\frac{\vec \alpha \cdot \vec \phi}2 \right) + \sin^2 \left(\frac{\vec \beta \cdot \vec \phi}2  \right) \right]\,, \label{compact-potential}
 \end{eqnarray}
where we write the two fields as a vector, $\vec \phi := [\phi_1, \phi_2]$, and assume the scales $\Lambda_i$ are identical (as might be arranged with a discrete symmetry\footnote{Next to calculational simplicity, this has the advantage that the direction with the largest possible displacement also is lightest \cite{Bachlechner1}.}). Higher-order corrections such as
 \begin{align}
   \sin^4  \left(\frac{\vec \alpha \cdot \vec \phi}2 \right) \,, \qquad  \sin^2 \left(\frac{\vec \alpha \cdot \vec \phi}2 \right) \sin^2  \left(\frac{\vec \beta \cdot \vec \phi}2 \right) \,,
\end{align}
are suppressed as they come with higher powers of $\Lambda$, which is required to be order of magnitudes below the Planck scale.
The two `alignment' vectors can be parametrized by
   \begin{align}
    & \vec\alpha = \lambda_1 \left[ \cos(\omega/2), \sin(\omega/2) \right] \,, \notag \\
    & \vec\beta = \lambda_2 \left[ \cos(\omega/2), - \sin(\omega/2) \right] \,. \label{vectors}
  \end{align}
Here we use the $SO(2)$ invariance of the kinetic term to absorb a irrelevant additional angle. The two quantities that determine the shape of this scalar potential are the ratio, $\lambda_2 / \lambda_1$, of the two lengths and the angle, $\omega$, between these two vectors. The resulting potential landscape is illustrated for two representative choices of parameters in figure 1.

\begin{figure}[h!]
\centering
\includegraphics[scale=0.35]{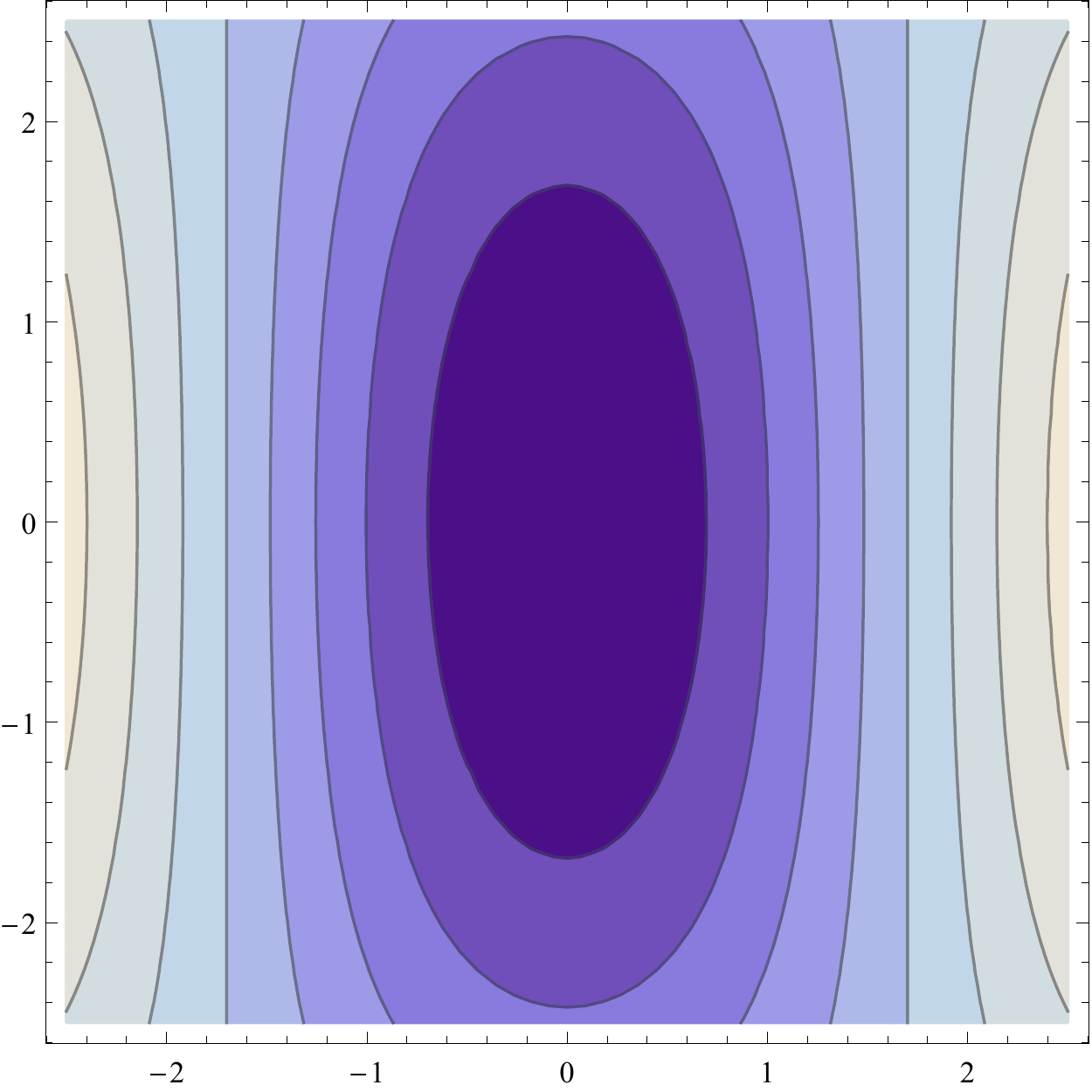}
\includegraphics[scale=0.35]{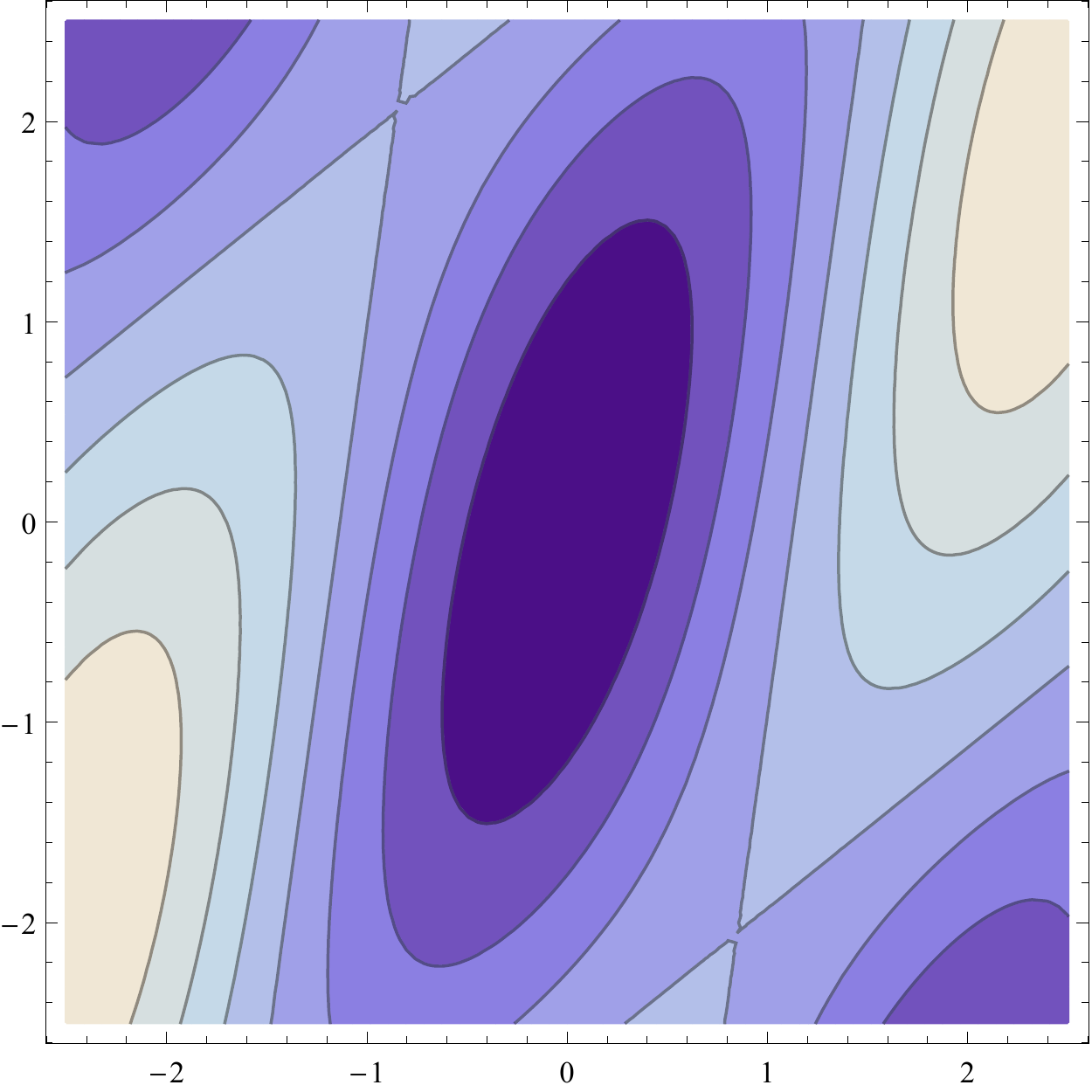}
\caption{The potential landscape \eqref{compact-potential} for $\omega = \pi/4$ and $\lambda_1 = \lambda_2$ (left panel) and $\lambda_1 = \lambda_2  /2 = 1$ (right panel). Darker colours indicate lower values of $V$.}
\end{figure}

The mass eigenvalues for this scalar potential are given by
 \begin{align}
  m_{\pm}^2/\Lambda^4 = \tfrac14 \left( \lambda_1^2 + \lambda_2^2 \pm \sqrt{\lambda_1^4+\lambda_2^4+2 \lambda_1^2 \lambda_2^2 \cos 2\omega} \right) \,,
 \end{align}
revealing a mass hierarchy for sufficiently small $\omega$:
 \begin{align}
 \frac{m_-^2}{\Lambda^4} = \frac{\lambda_1^2 \lambda_2^2 \, \omega^2}{2(\lambda_1^2+\lambda_2^2)} \,, \quad
 \frac{m_+^2}{\Lambda^4} = \frac{\lambda_1^2+\lambda_2^2}{2} \,.
 \end{align}
This hierarchy arises because when $\omega = 0$ the scalar potential only depends on one of the two axions, ensuring the orthogonal direction becomes flat. This is the lattice alignment mechanism of Kim, Nilles and Peloso\footnote{The generalization to more than two axions has been investigated by \cite{multi-axions}, while the embedding in string theory is the focus of e.g.~\cite{string-align}.} \cite{KNP}.

The connection to large $f$ is clearer if we study the same phenomenon in the lattice basis. In this case, the kinetic matrix reads
 \begin{align}
  K_{ij} = \frac{1}{\lambda_1^2 \lambda_2^2 \sin^2 \omega } \left(
 \begin{array}{cc} \lambda_2^2 & - \lambda_1 \lambda_2 \cos\omega \\
-  \lambda_1 \lambda_2 \cos\omega & \lambda_1^2 \end{array} \right)  \,. \label{compact-kinetic}
 \end{align}
In a lattice basis the fields individually range over the square interval $-\pi \le \theta_i \le \pi$, and using the metric (\ref{compact-kinetic}) the longest distance in this fundamental domain is
 \begin{align}
    \Delta\theta^2 = \frac{2 \pi (\lambda_1^2 + \lambda_2^2 \pm 2  \lambda_1 \lambda_2 \cos\omega)}{\lambda_1^2 \lambda_2^2 \sin^2\omega } \,,
 \end{align}
depending on whether one measures the length of a diagonal (-) or an anti-diagonal (+) in $\mathcal{M}_\theta$. As stressed in \cite{Bachlechner1}, at fixed eigenvalues, the field range is maximized by having the eigenvectors of $K$ align with the (anti-)diagonals. For the above matrix, such an alignment happens when $\lambda_1 = \lambda_2 = \lambda$, in which case the two eigenvalues of $K$ simplify to
 \begin{align}
  \frac{1}{2 \lambda^2 \sin^2(\omega/2)} \,, \quad  \frac{1}{2 \lambda^2 \cos^2(\omega/2)} \,.
 \end{align}
One of these diverges as $\omega \to 0$, showing how an $f_i$ can diverge without bound in the lattice-basis counterpart of lattice alignment.

This simple example highlights the following features:
\begin{itemize}
 \item
 In the kinetic basis it is the angle between the vectors in the scalar potential \eqref{compact-potential} that determines lattice alignment, and it is their length ratio that determines kinetic alignment. Perfect alignment corresponds to coinciding vectors with $\omega=0$ and $\lambda_1 = \lambda_2$.
 \item
 In the lattice basis it is the eigenvalues and eigenvectors of the kinetic matrix, $K_{ij}$, that determine lattice and kinetic alignment, respectively. Perfect alignment corresponds to an infinite eigenvalue in an (anti-)diagonal direction.
\end{itemize}

The inflationary analysis is particularly simple in the case of perfect kinetic alignment: $\lambda_1 = \lambda_2 = \lambda$. In this case, the potential (\ref{compact-potential}) can be written in the remarkably simple form
 \begin{align}
  V = \Lambda^4 \left[ 1 - \cos\left( \lambda \phi_1\cos\frac\omega2 \right) \cos \left(\lambda \phi_2 \sin\frac\omega2 \right) \right] \,.
 \end{align}
Moreover, the symmetries under the separate reflections, $\phi_1 \to -\phi_1$ and $\phi_2 \to -\phi_2$ in this case show that the single-field model appropriate to the bottom of the trough (for small $\omega$) is a simple truncation that sets $\phi_1 = 0$. In this truncation the effective potential for the inflaton, $\varphi := \phi_2$, along the trough's bottom becomes
 \begin{align}
  V_{\rm eff} \simeq \Lambda^4 \left[ 1 -  \cos\left(\varphi/ f_{\rm eff} \right) \right] \,, \label{eff-sin}
 \end{align}
with effective decay constant $1/f_{\rm eff} = \lambda \sin(\omega/2)$. This gives the $N$-flation Pythagorean gain (for $N=2$) of $\sqrt2$ when $\omega = \pi/2$, relative to the single-field case. In contrast, $f_{\rm eff}$ can become super-Planckian as $\omega$ decreases towards $0$. We have included the inflationary predictions of this model as a function of the effective decay constant in figure 2.

\begin{figure}[h!]
\centering
\includegraphics[scale=0.6]{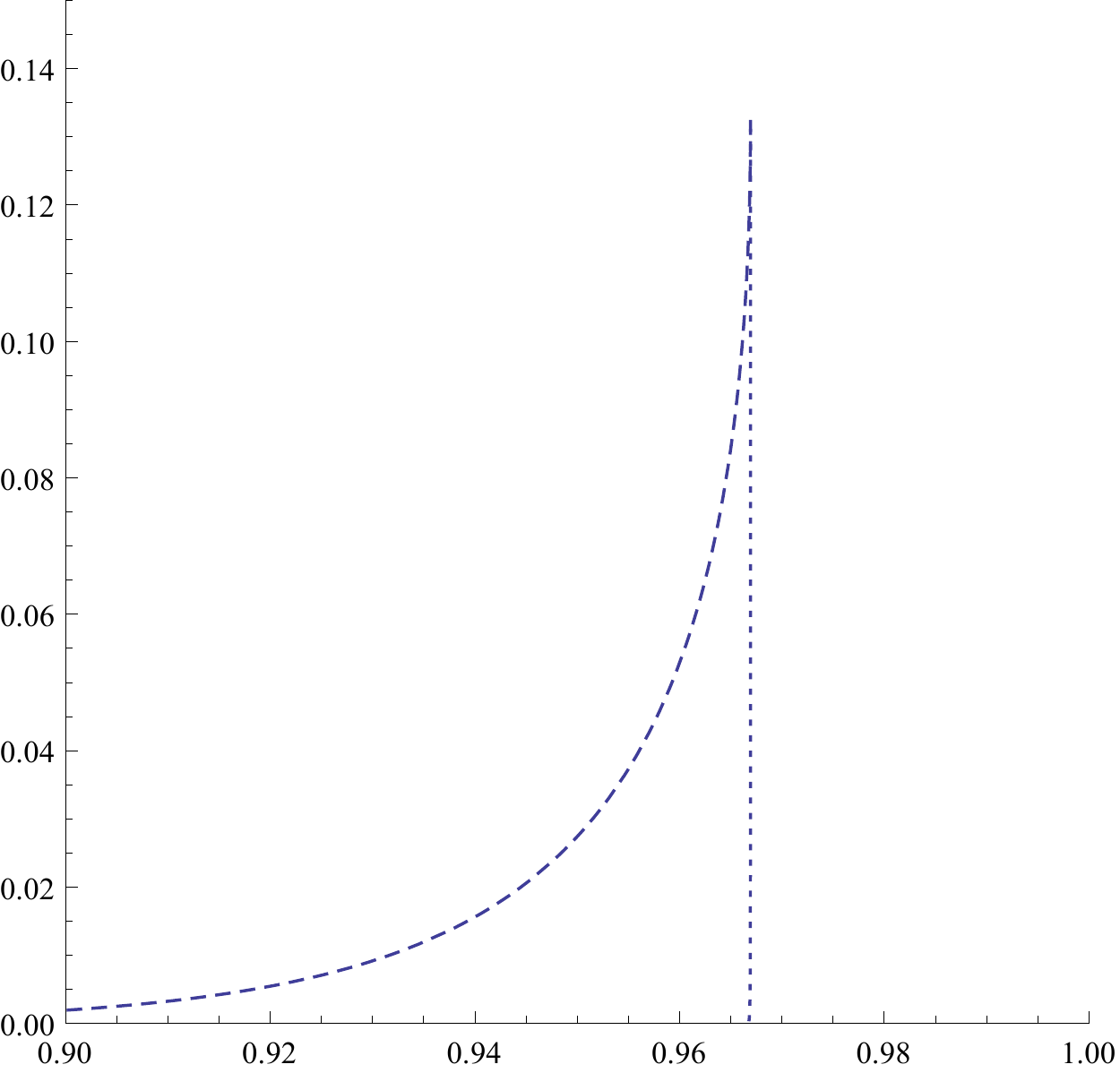}
\caption{The spectral index $n_s$ and tensor-to-scalar ratio $r$ (evaluated at 60 e-folds) for the effectively single-field models \eqref{eff-sin} (dashed line) and \eqref{eff-tanh} (dotted line) as a function of the effective decay constant.}
\end{figure}

\section{Noncompact case}

The above makes clear that nothing about the alignment mechanisms relies on the periodicity of the potential, and so can be applied equally well to non-axion models, such as noncompact pGBs. We now use this observation to two ends. First, we use it to obtain observably large tensor modes for large- and multi-field noncompact models (for which alignment is not required to obtain the basic slow-roll). Second, we use it to obtain slow-roll for multi-field noncompact models for which the corresponding single-field model would not itself inflate.

Before turning to more specific models we examine a simple case to which more complicated examples often reduce. Consider a positive-definite scalar potential (tuned to be minimized at zero) that can be written as a sum of squares of the form
 \begin{align}
  V =  U^2 ( \vec \alpha \cdot \vec \phi )  +   U^2 ( \vec \beta \cdot \vec \phi )  \,, \label{potential-tanh}
 \end{align}
where the real function $U(\theta)$ is at this point arbitrary. This contains in particular the axion example described earlier as the special case $U(\theta) = \Lambda^2 \sin(\theta/2)$. In the models of later interest it is assumed there are (usually different) points where each of $U$ and $U'$ vanish: $U(\theta_0) = U'(\theta_1) = 0$.

Once the function $U$ is fixed, alignment is fully specified by the two constant vectors $\vec \alpha$ and $\vec \beta$, that we parametrize as in \eqref{vectors}. Again, the two relevant properties are the relative angle $\omega$ and the ratio $\lambda_2 / \lambda_1$. When this is so we immediately have four stationary points for $V$ of (\ref{potential-tanh}):
\begin{enumerate}
\item[(I)] when $\vec\alpha \cdot \vec\phi = \vec \beta \cdot \vec\phi = \theta_0$;
\item[(II)] when $\vec\alpha \cdot \vec\phi = \vec \beta \cdot \vec\phi = \theta_1$;
\item[(III)] when $\vec\alpha \cdot \vec\phi = \theta_0$ and $\vec \beta \cdot \vec\phi = \theta_1$; and
\item[(IV)] when $\vec\alpha \cdot \vec\phi = \theta_1$ and $\vec \beta \cdot \vec\phi = \theta_0$.
\end{enumerate}
at which point the potential's second-derivative matrix becomes
\begin{enumerate}
\item[(I)] $V_{ij} = 2(\alpha_i \alpha_j+ \beta_i \beta_j)(U'_0)^2$;
\item[(II)] $V_{ij} = 2(\alpha_i \alpha_j+ \beta_i \beta_j) \,U_1 U_1''$;
\item[(III)] $V_{ij} = 2 \alpha_i \alpha_j (U'_0)^2 + 2 \beta_i \beta_j \,U_1 U_1''$;
\item[(IV)] $V_{ij} = 2 \alpha_i \alpha_j \,U_1 U_1'' + 2 \beta_i \beta_j (U'_0)^2$,
\end{enumerate}
where $U_0' := U'(\theta_0)$, $U_1 := U(\theta_1)$ and $U''_1 := U''(\theta_1)$. Clearly $V(\hbox{II}) >  V(\hbox{III}) \,,  V(\hbox{IV}) >  V(\hbox{I}) = 0$ and so generically (I) is a minimum and (II) is a maximum while (III) and (IV) are saddle points.

The alignment discussion applies just as for natural inflation, as this is independent of the details of the scalar potential. Thus the matrix $K_{ij}$ has the same crucial property that its eigenvectors align with the (anti-)diagonals provided the lengths of $\vec\alpha$ and $\vec \beta$ coincide. Given such kinetic alignment, the eigenvalues of $K$ can be tuned using the angle $\omega$. One eigenvalue diverges as $\omega$ approaches zero, ensuring an effective `decay constant' (whose physical interpretation depends on the function $U$) that blows up. In the kinetic basis, this corresponds to an almost flat direction that develops as the vectors $\vec\alpha$ and $\vec \beta$ start to align.

When the function $U$ is (anti-)symmetric, one can restrict $\omega$ to the first quadrant without loss of generality. Furthermore, given kinetic alignment, it is consistent to truncate to either $\phi_1$ or $\phi_2$. Given that $\omega$ is in the first quadrant, the inflationary trough lies in the direction of $\phi_2$: for $\omega \to 0$ this direction is exactly flat, while for $\omega = \pi/2$ the two terms in the scalar potential have opposite orientation and $\phi_1$ and $\phi_2$ have equal masses. Setting $\vec \phi = ( 0, \varphi)$, we obtain the effective potential along the trough bottom:
 \begin{align}
  V_{\rm eff} \simeq U^2 \left( \lambda \varphi \sin\frac\omega2 \right) \,,
 \end{align}
allowing the effective `decay constant' $1/( \lambda \sin\frac\omega2 )$ to be tuned in exactly the same way as for aligned natural inflation \cite{KNP}. Starting at $\omega = \pi/2$ one has an enhancement factor of $\sqrt{2}$, and this can be increased to arbitrary magnitude by alignment of the two vectors. Given a small-field expansion for the anti-symmetric function $U(\theta) = c_1 \theta + c_3 \theta^3 + \ldots$, for small enough $\omega$ the scalar potential is always dominated by the lowest-order quadratic terms, ensuring the resulting predictions are always those of quadratic inflation in this limit.

Another broad class of models with pGB inflatons comes from noncompact groups and leads to exponential potentials, of the broad form $V = V_0 - V_1 e^{-\phi/f} + \cdots$. It has been remarked that such an exponential series has two attractive properties: ($i$) the slow-roll conditions are {\em always} satisfied for sufficiently large $\phi$, regardless of the value of $f$; ($ii$) they can arise in compactifications where higher-dimensional lengths become the inflaton, and when they arise in this way the decay constant is naturally large: $f \simeq M_p$ \cite{BBarModuli,LFInf}. However large $f$ can still be useful since in these models it is a prerequisite for obtaining large primordial tensor modes. The next few examples explore several applications of the alignment mechanisms to such exponential potentials.

\subsection{Large tensor modes for exponential potentials}

We start with an example of a function that asymptotes to a plateau with exponentially suppressed fall-off for large $\phi$, but is anti-symmetric in $\phi \to - \phi$. One such a potential uses
\begin{equation}
 U(\theta)  = \frac{\Lambda^2}{\sqrt2} \; \tanh\theta \,,
\end{equation}
in which case the resulting hyperbolic scalar potential furnishes an analog of natural inflation. The corresponding single-field model was proposed in a different context \cite{conformal}, starting from motivations based on (super-)conformal symmetry breaking and are referred to as T-models.

Generalizing to two scalar fields as above gives
 \begin{align}
   V = \tfrac12 \Lambda^4 [ \tanh^2 (\vec \alpha \cdot \vec \phi) +  \tanh^2 (\vec \alpha \cdot \vec \beta ) ] \,,
  \end{align}
and the resulting potential landscape is illustrated in figure 3. With this example we have $\theta_0 = 0$ and $\theta_1 = \infty$ and so $U'_0 = U_1 = \frac{1}{\sqrt{2}} \, \Lambda^2$ and $U_1'' = 0$. Consequently $V_{ij}(\hbox{I}) = (\alpha_i \alpha_j + \beta_i \beta_j) \Lambda^4$, $V_{ij}(\hbox{II}) = 0$ while $V_{ij}(\hbox{III}) = \alpha_i \alpha_j \, \Lambda^4$ and $V_{ij}(\hbox{IV}) = \beta_i \beta_j \, \Lambda^4$.

\begin{figure}[h!]
\centering
\includegraphics[scale=0.35]{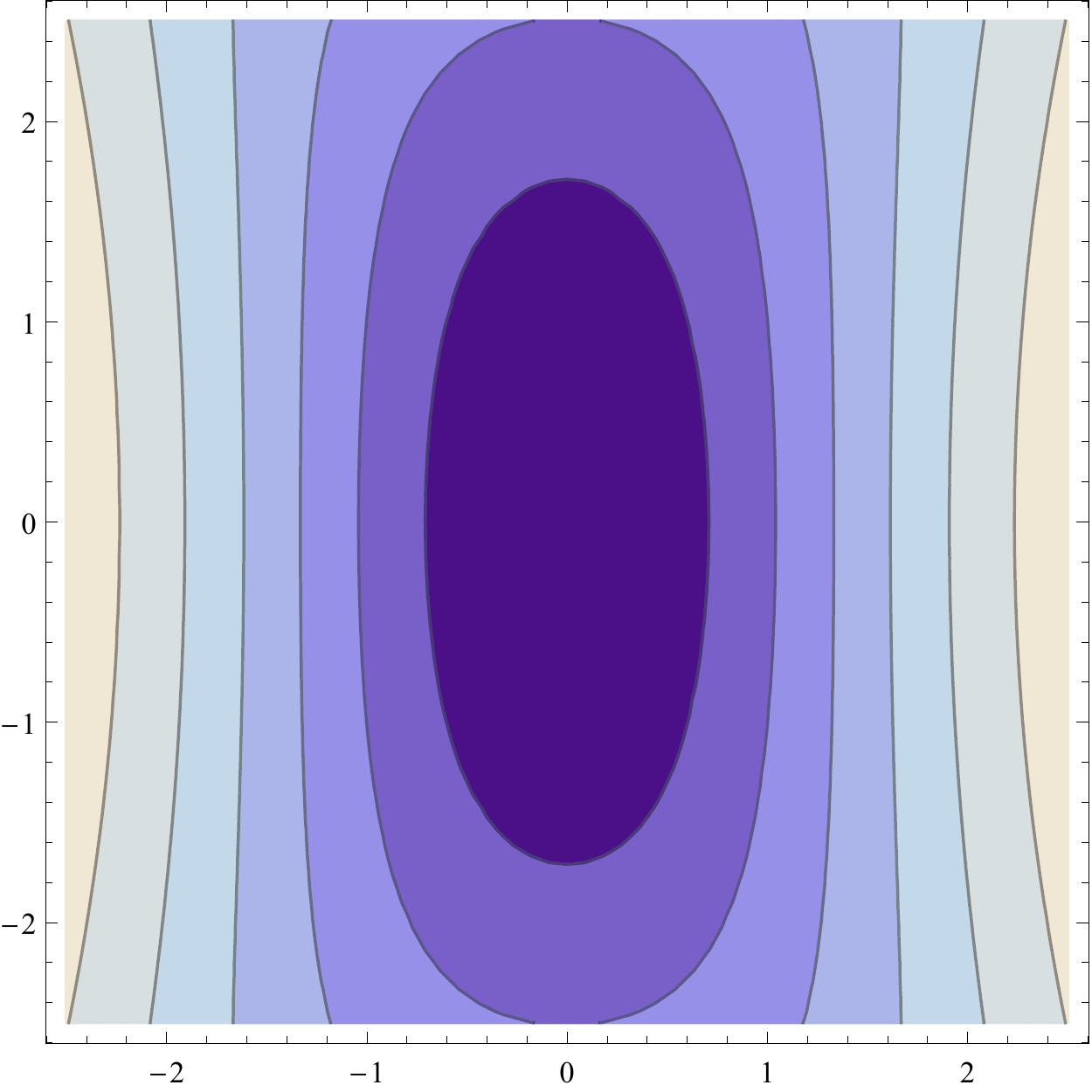}
\includegraphics[scale=0.35]{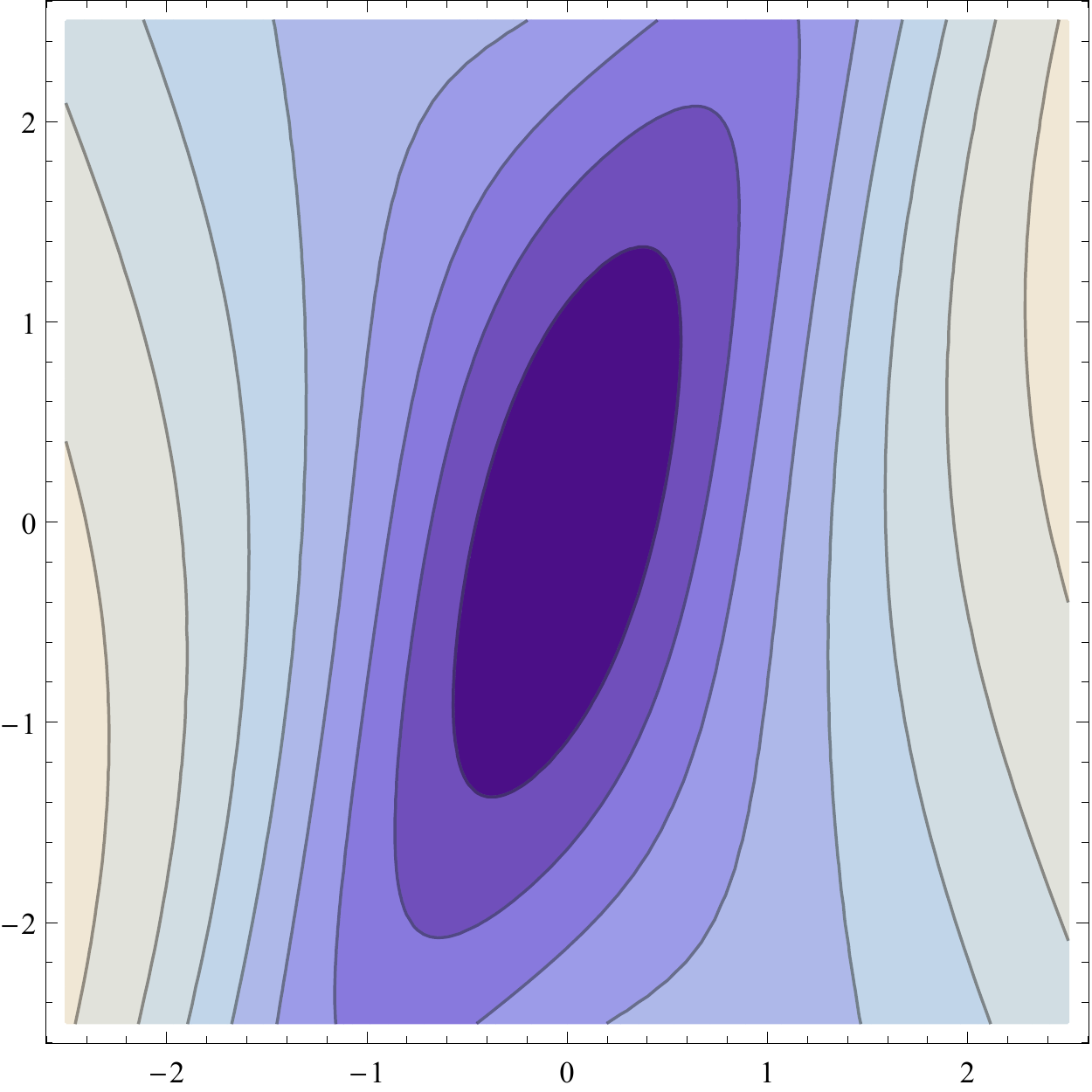}
\caption{The potential landscape for \eqref{potential-tanh} with $U(\theta) \propto \tanh \theta$ and for $\omega = \pi/4$ and $\lambda_1 = \lambda_2$ (left panel) and $\lambda_1 = \lambda_2  /2 = 1$ (right panel). Darker colours indicate lower values for the potential.}
\end{figure}

The inflationary analysis is particularly simple for maximal kinetic alignment, for which $\lambda_1 = \lambda_2 = \lambda$ and we can take $\omega$ in the first quadrant. In this case one can again truncate to equate the inflaton with $\phi_2$, leading to the following effective inflaton potential
 \begin{eqnarray}
 V_{\rm eff} &\simeq& \Lambda^4 \tanh^2 \left( \lambda \varphi \sin\frac\omega2 \right) \nonumber\\
 &\simeq& \Lambda^4 \left( 1 - 2 \, e^{-2\varphi/f_{\rm eff}} + \cdots \right)^2 \,, \label{eff-tanh}
 \end{eqnarray}
where the last approximate equality applies when $\varphi \gsim f_{\rm eff}$, with $1/f_{\rm eff} := \lambda \sin(\omega/2)$. The model's inflationary predictions are now easy to infer. For generic $\omega$ inflation takes place out near the plateau at $\varphi \gg f_{\rm eff}$, where the slow-roll conditions are satisfied because $e^{-\varphi/f_{\rm eff}} \ll 1$. Hence its properties are determined by the first, exponential fall-off term. This leads to a spectral index and tensor-to-scalar ratio that are given by the standard result for exponential potentials \cite{conformal},
 \begin{align}
  n_s = 1- \frac{2}{N} \,, \quad r = \frac{2}{f_{eff}^2 N^2} \simeq \cO[( n_s -1)^2]\,, \label{Staro}
  \end{align}
at lowest order in $1/N$, and in particular $r \lsim \mathcal{O}(0.001)$ which could eventually be observable, but not yet. By contrast, when lattice alignment occurs and the angle between the two vectors approaches zero, inflation takes place at smaller values of $\varphi/f_{\rm eff}$, for which $M_p \lsim \varphi \ll f_{\rm eff}$ and so the long trough is effectively quadratic in $\varphi$. This then leads to standard $m^2\varphi^2$ predictions \cite{PhiSqInf}
 \begin{align}
  n_s = 1- \frac{2}{N} \,, \quad r = \frac{8}{N} = 4(1 - n_s)\,, \label{Quadr}
  \end{align}
modulo subleading corrections. In particular, $r \simeq 0.16$ is much easier to observe in this regime. Again, the inflationary predictions of this model as a function of the effective decay constant can be found in figure 2. Closely related interpolations between the above two predictions have been pointed out in \cite{DoubleAttractors}.

A similar argument also applies more generally to more generic two-field exponential potentials. For instance consider a class of positive potentials, $V \ge 0$, with minima tuned to lie at $V = 0$ (similar in spirit to the tuning done for axion potentials), but with a large-field asymptotic form
\begin{equation}
 V = V_0 - V_1 e^{- \theta_1} - V_2 e^{- \theta_2} + \cdots \,,
\end{equation}
where $\theta_i$ is large enough to neglect terms represented by the ellipses --- involving at least two powers of $e^{-\theta_i}$. Suppose also that $V_0$ happens to be much smaller than all of the other $V_i$, so that the $V = 0$ minima also lie in the large-field regime. In this case we may write
\begin{equation}
 V \simeq \frac{V_0}2 \left( 1 - \frac{V_1}{V_0} \; e^{-\theta_1} \right)^2 + \frac{V_0}2 \left( 1 - \frac{V_2}{V_0} \; e^{-\theta_2} \right)^2 + \cdots\,,
\end{equation}
which has minima when $e^{-\theta_i} = V_0/V_i \ll 1$. If $V_1 = V_2$ (as might be arranged through a symmetry), then (writing $V_0 = \Lambda^4$) this has the approximate form of (\ref{potential-tanh}), with
\begin{equation}
 U(\theta) = \frac{\Lambda^2}{\sqrt2} \left( 1 - A \, e^{-\theta}  \right) \,.
\end{equation}
Because $U(\theta)$ is not (anti-)symmetric, the trough obtained by alignment for small $\omega$ is not along the $\phi_2$ axis, and so the effective potential along the minimum is obtained using methods adapted to curved troughs \cite{TroughInf,AA2} rather than simply by truncation. Simpler, reflection-symmetric, exponential potentials can also happen, such as for non-minimally coupled scenarios like Higgs inflation \cite{HI}, when the exponential asymptotics are functions of $e^{-|\theta|}$ rather than $e^{-\theta}$ \cite{UA}.

\subsection{Flat directions from alignment}

In the previous noncompact examples alignment is unnecessary for slow-roll and instead is used to raise the inflationary scale and so enhance the production of primordial tensor modes. In this section we instead examine an example with exponential potentials for which alignment is responsible for the slow-roll itself. We do so to present a cartoon of how inflation using extra-dimensional moduli (such as the radius, $r_i$, of an extra-dimensional cycle) can arise in more complicated examples \cite{KMI}.

Extra-dimensional moduli can generically arise with exponential potentials because (for small curvatures) their energy usually arises in a derivative expansion, as powers of $1/r_i$. However the kinetic terms such fields inherit from the higher-dimensional Einstein action can be of the form $M_p^2 (\partial r_i)^2/r_i^2$, making the canonical field $\theta_i \sim \ln r_i$. Their origin from the Einstein action also ensures $\lambda_i \sim 1/M_p$ for these modes.

What is hard to get with moduli is a constant energy density ({\em i.e.} the term $V_0$ in the previous example) as $r_i \to \infty$, and so we instead explore in this section the choice
\be \label{expU}
 U(\theta) = \frac{\Lambda^2}{\sqrt2} \left( e^{-\theta} - A e^{-2\theta} \right) \,,
\ee
in the two-modulus potential (\ref{potential-tanh}). We again assume $A \gg 1$ so that the potential's minimum occurs at large fields: $e^{-\theta_0} = 1/A \ll 1$. In this case there is no slow-roll for generic values of $\omega$ since there is no asymptotic plateau out at large fields.

With this choice we have $e^{\theta_0} = A$ and $e^{\theta_1} = 2A$ and so $4\sqrt2 \; U_1 = -2\sqrt2\; U_1'' = \sqrt2 \; U_0' = \Lambda^2/A$, and the resulting potential landscape is plotted in figure 4. This shows the minimum at point (I) that is separated from the asymptotic falloff to zero at large $\theta_i$ by the maximum at point (II) as well as the two saddle points (III) and (IV). As $\omega$ goes to zero the trough running from the minimum to the saddle points elongates, eventually allowing inflationary solutions that roll down towards the minimum from the saddle points.

\begin{figure}[h!]
\centering
\includegraphics[scale=0.35]{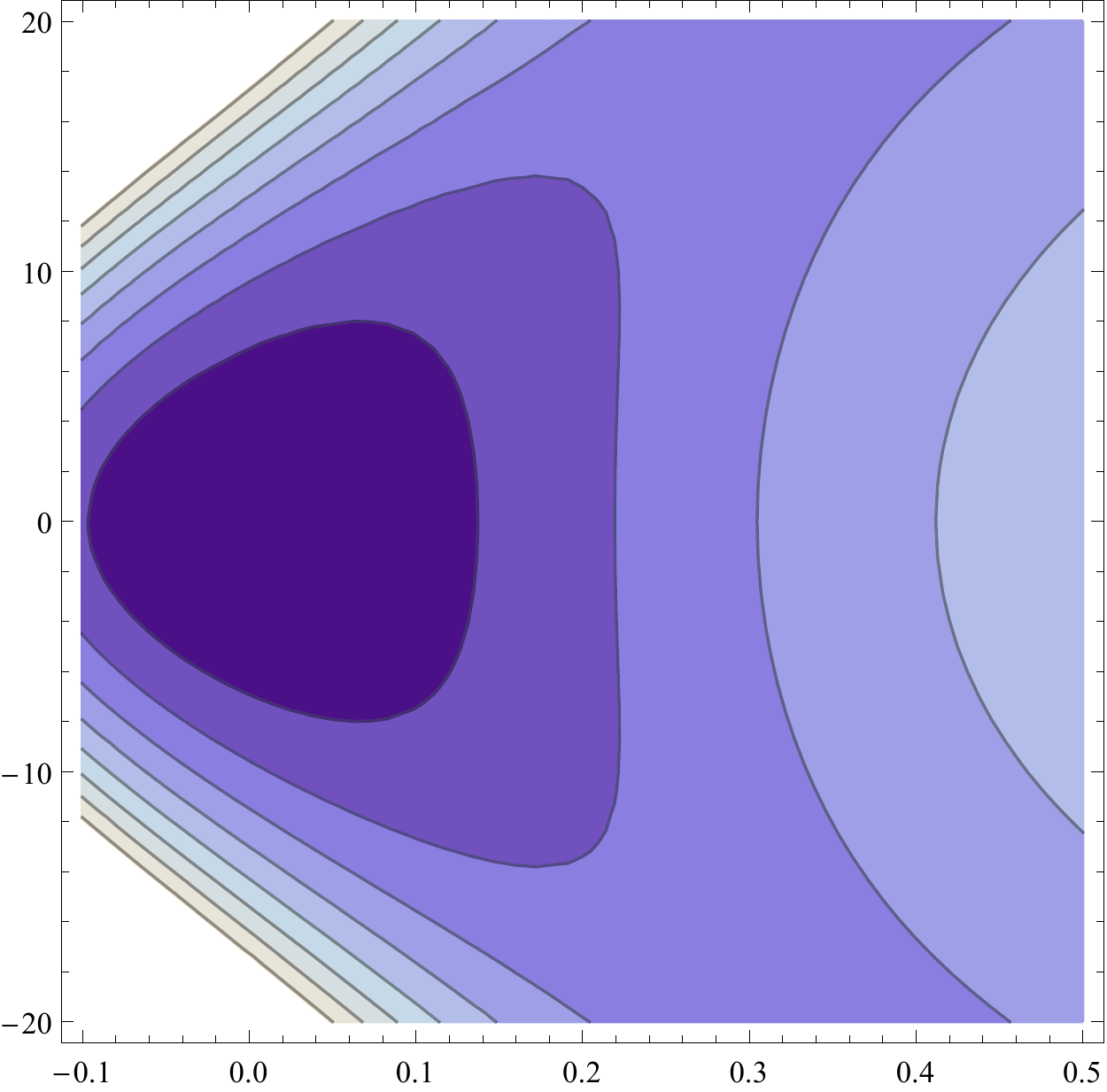}
\includegraphics[scale=0.35]{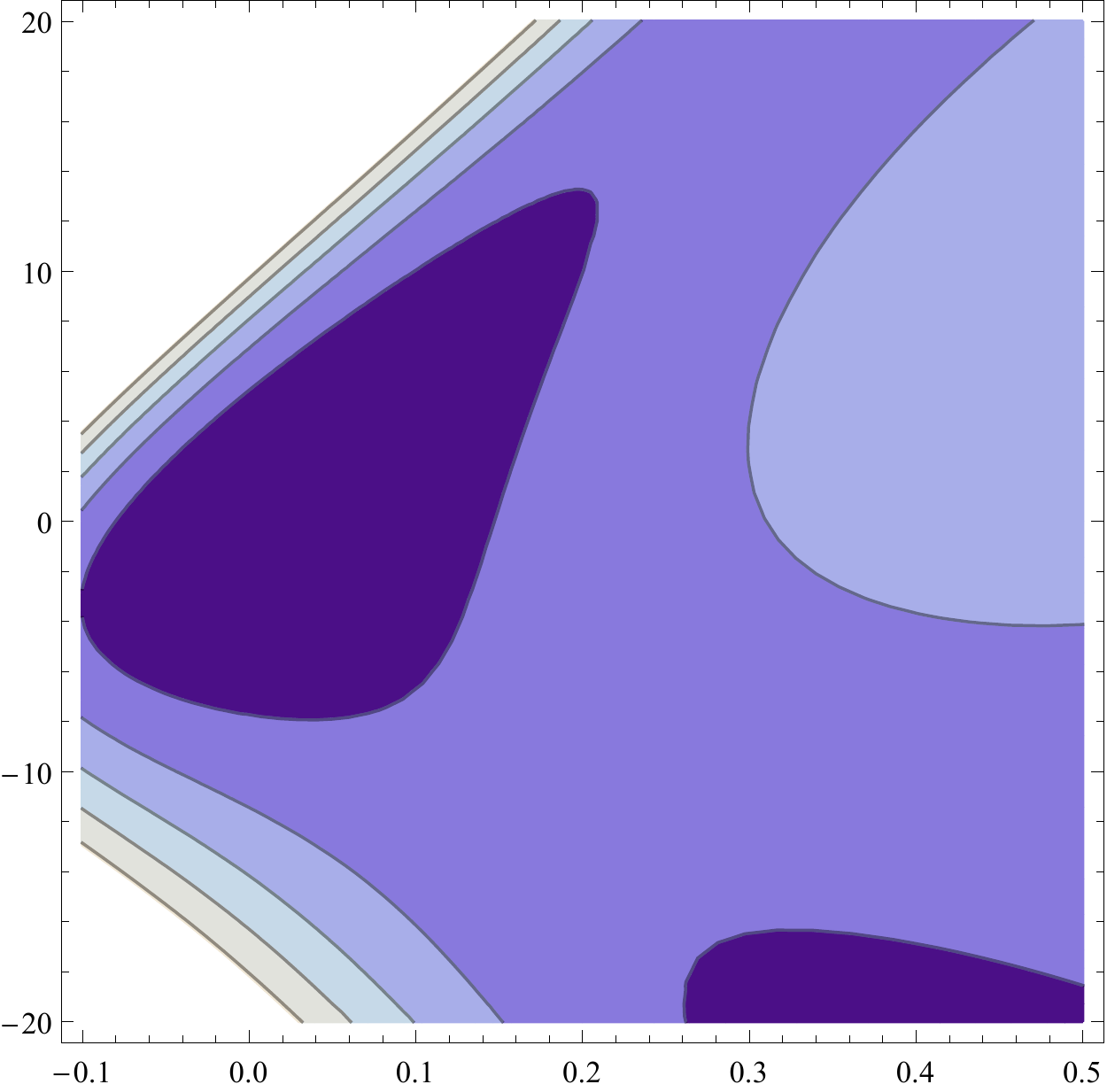}
\caption{The potential landscape for the choice (\ref{expU}) with $\omega = \pi/100$ and $A=\lambda_1 = \lambda_2=1$ (left panel) and $A=\lambda_1 = \lambda_2  /2 = 1$ (right panel). Darker colours indicate lower values of $V$.}
\end{figure}

Although the potential along the trough is not in general well-captured by truncation onto $\phi_2$ it does become effectively quadratic (and so reproduce the inflationary predictions of quadratic potentials) in the small-$\omega$ limit.

\section{Discussion}

In this note we have highlighted the versatility of alignment for inflation. Two mechanisms have been discussed to facilitate inflation based on a number of compact psGs, i.e.~natural inflation: kinetic alignment, extending the possible field range, as well as lattice alignment, increasing the effective decay constant. We have elucidated the relation between these two in the simple case of two axions. Moreover, we have applied the same alignment techniques to a different class of inflationary models based on non-compact pGBs. While our focus has been on two specific examples, based on hyperbolic (in section III.A) and expontential (in III.B) functions, it is clear that the applicability of alignment in inflation far exceeds these specific models.

\section*{Acknowledgements}

We thank M. Cicoli, R. Diener, S. de Alwys, L. McAllister, F. Quevedo and M. Williams for useful discussions. CB's research was supported in part by funds from the Natural Sciences and Engineering Research Council (NSERC) of Canada. Research at the Perimeter Institute is supported in part by the Government of Canada through Industry Canada, and by the Province of Ontario through the Ministry of Research and Information (MRI).

\end{document}